\documentclass[]{spie}  

\usepackage{amsmath,amsfonts,amssymb}
\usepackage{graphicx}
\usepackage{hyperref}
\hypersetup{colorlinks=true, allcolors=blue}

\usepackage{listings}

\usepackage[frozencache=true,cachedir=minted-cache]{minted} 
\usepackage{xcolor} 
\definecolor{LightGray}{gray}{0.9}
\setminted[python]{
baselinestretch=1.2,
bgcolor=LightGray,
fontsize=\footnotesize,
linenos
}

\title{Towards virtual access of adaptive optics telemetry data}

\author[a,f]{Tiago Gomes}
\author[b,f]{Carlos Correia}
\author[e]{Lisa Bardou}
\author[c]{Olivier Beltramo-Martin}
\author[c,d]{Thierry Fusco}
\author[g]{Caroline Kulcsár}
\author[e]{Timothy Morris}
\author[h,f]{Nuno Morujão}
\author[c]{Benoît Neichel}
\author[e]{James Osborn}
\author[a,f]{Paulo Garcia}

\affil[a]{Faculdade de Engenharia da Universidade do Porto, Rua Dr. Roberto Frias, s/n, 4200-465 Porto, Portugal}
\affil[b]{Space ODT - Optical Deblurring Technologies, Porto, Portugal}
\affil[c]{Aix Marseille Univ, CNRS, CNES, LAM, Marseille, France}
\affil[d]{DOTA, ONERA, Université Paris Saclay, F-91123 Palaiseau, France}
\affil[e]{Durham University (United Kingdom)}
\affil[f]{Center for Astrophysics and Gravitation, Instituto Superior Técnico, Av. Rovisco Pais 1, 1049-001 Lisboa, Portugal}
\affil[g]{Université Paris-Saclay, Institut d'Optique Graduate School, CNRS, Laboratoire Charles Fabry, 91127 Palaiseau, France}
\affil[h]{Departamento de Física e Astronomia, Faculdade de Ciências da Universidade do Porto, Rua do Campo Alegre s/n, 4169-007 Porto, Portugal}

\begin{document} 
\maketitle

\begin{abstract}
Large amounts of Adaptive-Optics (AO) control loop data and telemetry are currently inaccessible to end-users. Broadening access to those data has the potential to change the AO landscape on many fronts, addressing several use-cases such as derivation of the system’s PSF, turbulence characterisation and optimisation of system control. We address one of the biggest obstacles to sharing these data: the lack of standardisation, which hinders access. We propose an object-oriented Python package for AO telemetry, whose data model abstracts the user from an underlining archive-ready data exchange standard based on the Flexible Image Transport System (FITS). Its design supports data from a wide range of existing and future AO systems, either in raw format or abstracted from actual instrument details. We exemplify its usage with data from active AO systems on 10m-class observatories, of which two are currently supported (AOF and Keck), with plans for more.
\end{abstract}


\section{INTRODUCTION}
\label{sec:intro}
Despite the increasing adoption of Adaptive Optics (AO) systems in astronomy, accessing and interpreting their telemetry data remains a real challenge. We use the term ``telemetry'' to represent AO internal signals such as wavefront sensor measurements, deformable mirror commands and several other reconstruction and control matrices and parameters. AO telemetry data has the potential to be used in many applications of high scientific interest, e.g., instrumentation research, deriving the state of the atmospheric turbulence during observations\cite{Vidal-a-10, guesalaga2014using, Martin2016WHT, Jolissaint:18, Laidlaw2019Automated, andrade2019estimation}, deriving the point spread function\cite{1997JOSAA..14.3057V, 2006A&A...457..359G, gilles2012simulation, 2019MNRAS.487.5450B, 2020A&A...635A.208F, 2020MNRAS.494..775B}, performing system performance estimation or runtime optimization\cite{sivo2014first, petit2014sphere, sinquin2020sky}. Contradictorily, this type of data has been historically seen as ``engineering data'' and thus third-party access is disregarded by most data-producing systems\cite{hirst2020telemetry}. While most of these systems already record large amounts of telemetry for internal use (including one-off instrument commissioning and regular calibration), these data are usually saved in private archives that are out of reach to the end-user, or it is simply not kept in the long-term. Even in cases where third-parties can access telemetry datasets, their documentation tends to be poor or non-existent. This complicates the task of interpreting and analysing them, given that the data points being shared and how they are structured usually varies significantly between systems (and sometimes even between datasets generated by the same system). In sum, the landscape is dominated by a complete lack of uniformity in the data, which ends up being one of the biggest obstacles to data access.

Given the extensive use-cases for AO telemetry data, broadening its access could change the AO and astronomy landscape on many fronts. Accordingly, many systems are starting to show interest in saving and sharing such data. One of the earliest examples of public archiving of AO telemetry was carried out by the CANARY project\cite{canary2008}, who made its open loop data for the sodium laser guide star experiment\cite{bardou2021canary} available on the ESO Science Archive\cite{canary_archive}, with a data format that is extensively detailed in an accompanying manual\cite{canary_manual}. The Gemini Observatory is planning on publishing one of the first large scale archives of telemetry data\cite{hirst2020telemetry} (mostly composed on wavefront slopes and DM commands), through the Gemini Observatory Archive\cite{hirst2016archive}. However, publishing AO telemetry data currently requires significant effort from the responsible teams, as they are forced to design their own purpose-made data format, and make it available along with proper documentation for their prospective users. While such formats represent a significant step forward in data accessibility, given that they are tailored specifically for one instrument/observatory, they typically make no attempts at generalising support for multiple systems. In fact, to our knowledge, there has been no major attempt in the AO community to establish a consensual data format for AO telemetry to date.

We believe that access to AO telemetry data can only become ubiquitous by having the bulk of the AO community agree on a common way of sharing their data, that is, agreeing on a data exchange standard. A data exchange standard essentially defines a set of attributes/fields, their meanings, the way they relate to other data and the way they are stored in a shareable format. An ideal data exchange standard for AO telemetry would be able to package all relevant data in an unambiguous manner that is consistent across as many systems as possible, allowing data to be easily shared between data-producing systems and their end-users. It should also be structured in a generalised way that abstracts the user from observatory or instrument-specific details, without losing access to important information. While some of its fields should be mandatory in order to ensure the uniformity of data access, a significant portion should be optional, granting its flexibility to adapt to the irregularities of the telemetry being gathered by the different real-life instruments. Finally, it should be designed in a way that allows it to be continuously expanded to meet user needs and advancements in the AO field.

With all this in mind, we are developing AOT (Adaptive Optics Telemetry), a data exchange standard for AO telemetry data. This standard draws some inspiration from OIFITS\cite{pauls_data_2004, pauls_data_2005, duvert_oifits_2017}, which was created as a result of similar challenges that the interferometry community has faced. AOT is built on top of the Flexible Image Transport System (FITS)\cite{wells_fits-flexible_1979}, meaning that AOT files fully comply with the latest FITS standard\cite{chiappetti_definition_2018}, on top of which we define AOT-specific keywords and extensions. Specifically, we group data from different parts of the system with a set of 10 FITS binary tables, followed by as many image extensions as necessary to specify multi-dimensional data related to the system. While FITS is not an ideal standard for storing object-oriented data, building on top of FITS ensures significant compatibility with many of the existing tools, since any tool that fully supports the latest FITS standard will also be able to read AOT files, without requiring any modifications. A full discussion and exact specification of the first version of the AOT standard will be published separately at a later date. In this paper, we will be focused on the supporting mechanisms and tools that we have built for AOT.

The rate of adoption of a data exchange standard is its biggest metric of success, as low adoption would turn it into just another incompatible way of storing data, losing its intended purpose. Therefore, we are committed to alleviate anything that may prove to be a barrier to adoption. Specifically, we have developed \textit{aotpy} (Adaptive Optics Telemetry for Python), an object-oriented Python\cite{guido2009python} package that supports this data exchange standard. The goal of this package is to facilitate reading, editing and writing of AOT files, by abstracting the user from the actual file handling and its structure. It does so by implementing a data model through which the user can interact with all relevant data. We also provide preliminary ``Translator'' methods, currently for two AO systems on 10m-class observatories. These methods essentially act as interfaces between non-standard data currently being produced by these systems and our standard data model, allowing the user to work with supported legacy data via \textit{aotpy} and in turn create standard AOT files for it. This feature is important as it bridges the gap for legacy systems that are unlikely to adopt a newer standard.

This project is being developed in the context of OPTICON–RadioNet Pilot, an European collaboration through which we aim to promote broad discussion and consensus within the AO community. We plan on making available on the ESO Science Archive a set of AOT files for existing ESO systems; preliminary talks have been established with responsible teams, to ensure that no showstoppers exist.

In this paper we will be going through the implementation of the \textit{aotpy} package, code demonstrations, examples of its use-cases and the work that we have currently planned for the future.

\section{PYTHON PACKAGE}
\label{sec:Package}
\textit{aotpy} is an object-oriented Python package that defines a set of classes and variables which can hold all data that may be shared through an AOT file, as well as interfaces for writing and reading such files. In the future, this package will also provide tools to explore data in AOT files interactively.

While currently this package only provides support for FITS files, it was designed with the intention of enabling an easy expansion to other file types, as long as these can still be represented by the \textit{aotpy} data model.

\subsection{Data Model and Classes}
 The classes in \textit{aotpy} serve as abstractions of physical objects, their environment or the way they interact and behave in a AO system. Similar standpoints have been adopted in AO simulation tools such as OOMAO\cite{conan2014object}. They are implemented as Python Data Classes and provide access to a set of related AO data, which aims to be generalised enough to uniformly support distinct AO systems (classical with single natural or laser guide stars, multi-wavefront sensor systems, tomographic, etc.). For this purpose, common object-oriented techniques such as inheritance and polymorphism are employed.

\begin{figure}[ht]
   \begin{center}
   \includegraphics[width=\textwidth]{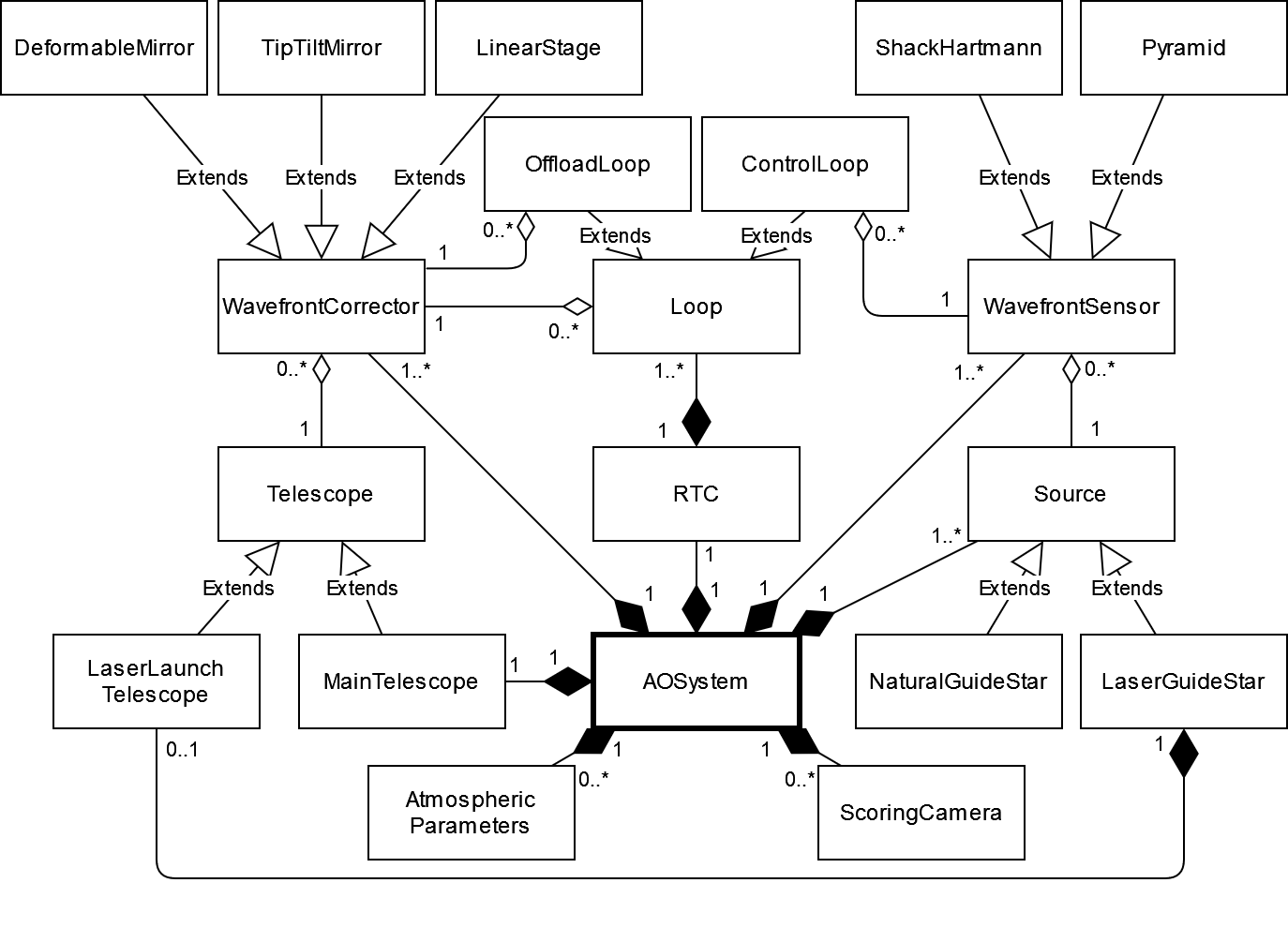}
   \end{center}
   \caption{\label{fig:UMLClassDiagram} 
    A language-agnostic UML Class Diagram that provides an overview of the data model and classes in \textit{aotpy}.}
\end{figure}

An overview of the data model implemented by \textit{aotpy} is provided in Fig.~\ref{fig:UMLClassDiagram}. Each AOT file can be fully represented by a single instance of \textit{AOSystem}, which contains telemetry from a single observation, as well as other data that may contextualise and support the analysis and interpretation of this telemetry.

A general description of all classes in \textit{aotpy} is provided in the following sections. Note that most classes have a \textit{name} variable, which is a unique user-defined identifier (that is, no other objects of the same type may have the exact same name). This is so that, even in scenarios where direct references to objects are not possible (for example, within a FITS file), these objects may still be uniquely referenced via their names.

\subsubsection{AOSystem}
\begin{figure}[htbp]
   \begin{center}
   \includegraphics[scale=0.27]{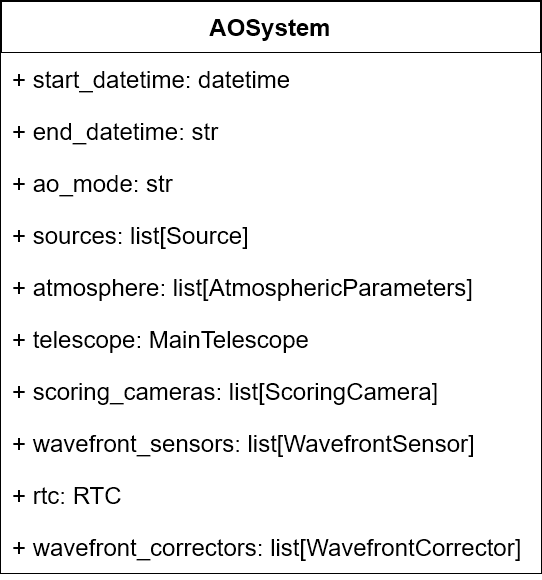}
   \end{center}
   \caption{\label{fig:ao_system} 
    UML Class Diagram for the \textit{AOSystem} class.}
\end{figure}

This is the central class in \textit{aotpy}. Its purpose is to provide general information about the observation, as well as references to all the different objects that detail the system (specifically, it references \textit{Source}, \textit{AtmosphericParameters}, \textit{MainTelescope}, \textit{ScoringCamera}, \textit{WavefrontSensor}, \textit{RTC} and \textit{WavefrontCorrector} objects). A diagram defining this class can be found in Fig.~\ref{fig:ao_system}.

\subsubsection{Image}
This class provides a generalised interface for multi-dimensional arrays of values. The array itself is stored as a NumPy\cite{harris2020array} \textit{ndarray}, but some metadata (such as units of measurement) can also be associated with an \textit{Image} object. 

Other classes in \textit{aotpy} contain references to \textit{Image} objects whenever some data cannot be represented as a single value or as a one-dimensional list of values. The same \textit{Image} object may be referenced by multiple objects in the system, avoiding data duplication.

\begin{figure}[htbp]
   \begin{center}
   \includegraphics[scale=0.27]{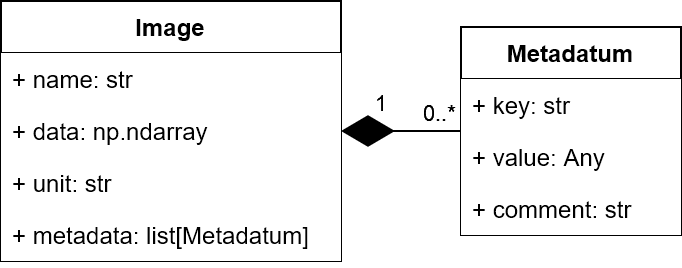}
   \end{center}
   \caption{\label{fig:image} 
    UML Class Diagram for the \textit{Image} and \textit{Metadatum} classes.}
\end{figure}

A representation of this class can be seen in Fig.~\ref{fig:image}.

\subsubsection{AtmosphericParameters}
\begin{figure}[htbp]
   \begin{center}
   \includegraphics[scale=0.25]{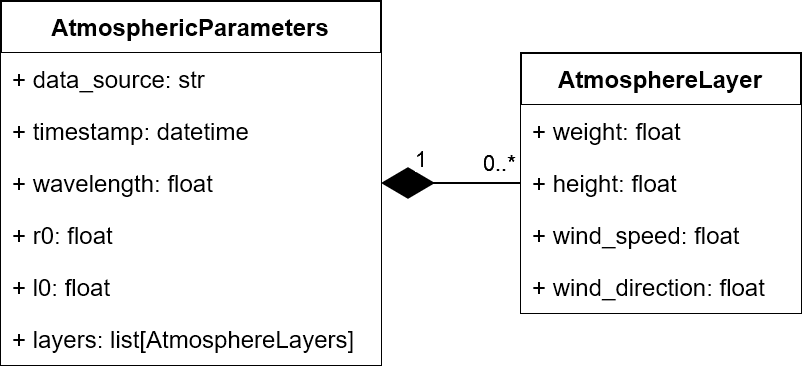}
   \end{center}
   \caption{\label{fig:atmosphere} 
    UML Class Diagram for the \textit{AtmosphericParameters} and \textit{AtmosphereLayer} classes.}
\end{figure}

Atmospheric data that may be relevant to the observation in question, mainly in regards to astronomical seeing and turbulence profiles, can be stored in \textit{AtmosphericParameters} instances. The source of such data, as well as the moment when it was recorded, is coupled with the data itself. This is shown in Fig.~\ref{fig:atmosphere}.

\subsubsection{Source}
\begin{figure}[htbp]
   \begin{center}
   \includegraphics[scale=0.27]{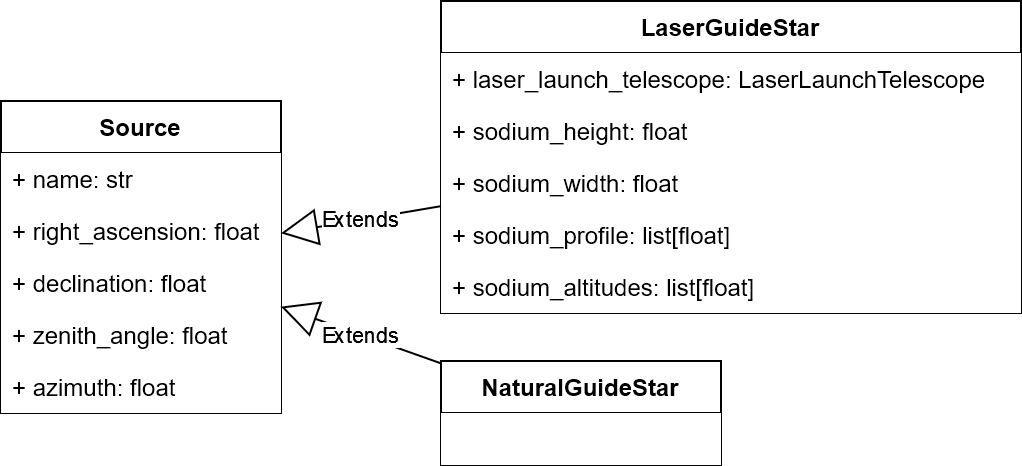}
   \end{center}
   \caption{\label{fig:source} 
    UML Class Diagram for the \textit{Source} class and its subclasses \textit{LaserGuideStar} and \textit{NaturalGuideStar}.}
\end{figure}

In \textit{aotpy}, \textit{Source} objects hold data related to the positioning of a light source. They shall be instantiated as either a \textit{NaturalGuideStar} or a \textit{LaserGuideStar}. The latter may contain further data related to the laser launch telescope used to create the source, and the sodium layer being targeted by it, as seen in Fig.~\ref{fig:source}.

\subsubsection{Telescope}
\textit{Telescope} objects contain data about a telescope's physical characteristics and configuration at the time of observation. These objects should be instantiated as \textit{MainTelescope}, if they contain data regarding the telescope at which the observation itself is performed, or \textit{LaserLaunchTelescope} if their data relates to the telescopes used to create artificial guide stars.

\begin{figure}[htbp]
   \begin{center}
   \includegraphics[scale=0.27]{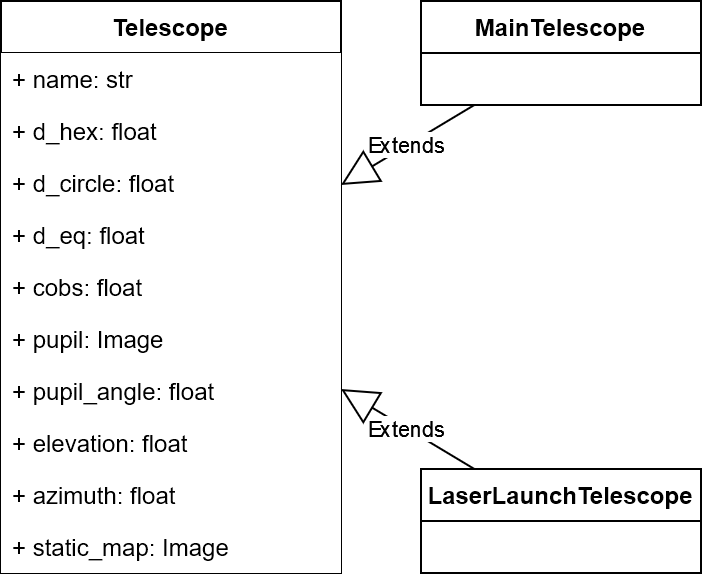}
   \end{center}
   \caption{\label{fig:telescope} 
    UML Class Diagram for the \textit{Telescope} class and its subclasses \textit{MainTelescope} and \textit{LaserLaunchTelescope}.}
\end{figure}

Each \textit{AOSystem} must only contain one \textit{MainTelescope} object (but as many \textit{LaserLaunchTelescope} objects as necessary). The contents of these classes are represented in Fig.~\ref{fig:telescope}.

\subsubsection{Detector}
\begin{figure}[htbp]
   \begin{center}
   \includegraphics[scale=0.27]{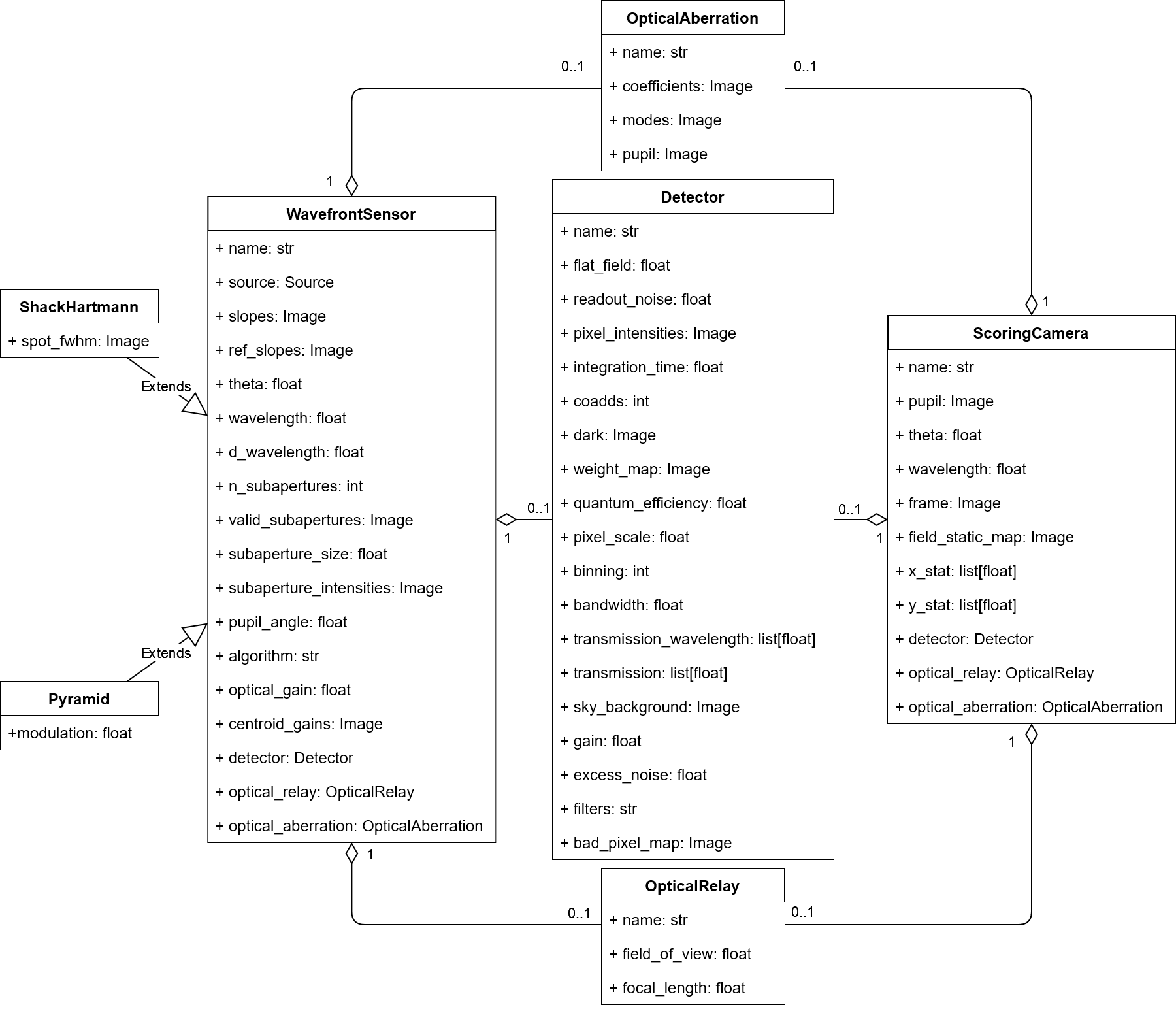}
   \end{center}
   \caption{\label{fig:optical_sensor} 
    UML Class Diagram for all the classes related to optical sensors. This includes \textit{Detector}, \textit{OpticalRelay}, \textit{OpticalAberration}, \textit{ScoringCamera}, \textit{WavefrontSensor} and its subclasses \textit{ShackHartmann} and \textit{Pyramid}}
\end{figure}

\textit{Detector} objects are an integral part of objects like \textit{ScoringCamera} and \textit{WavefrontSensor}. A \textit{Detector} object may hold a wide set of data related to the physical characteristics and configurations of a detector, as well as the actual pixels recorded by it. A detailed representation of its contents is provided in Fig.~\ref{fig:optical_sensor}.

\subsubsection{OpticalRelay}
An instance of this class contains data related to the physical configuration of the optical assembly, which may be present in \textit{ScoringCamera} and \textit{WavefrontSensor} objects. This is a rather simple class, which is represented in Fig.~\ref{fig:optical_sensor}.

\subsubsection{OpticalAberration}
This class provides variables that allow a user to describe optical aberrations. In \textit{aotpy} these may be associated to \textit{ScoringCamera} and \textit{WavefrontSensor} objects (as represented in Fig.~\ref{fig:optical_sensor}) but also to \textit{WavefrontCorrector} objects.

\subsubsection{ScoringCamera}
A \textit{ScoringCamera} object may contain some data about the characteristics of the camera, as well as references to related \textit{Detector}, \textit{OpticalRelay} and \textit{OpticalAberration} objects. However, it is important to note that the science data itself is not included here, or anywhere else in the package or the AOT standard, given that the focus is in handling telemetry data (usually, science data is already accessible separately). Fig.~\ref{fig:optical_sensor} details this class as well as its relation with other classes.

\subsubsection{WavefrontSensor}
A \textit{WavefrontSensor} object may contain a wide set of data related to its subapertures, the detected wavefront (such as gradients and intensities) and the algorithms and gain values involved. It can also provide references to related \textit{Detector}, \textit{OpticalRelay} and \textit{OpticalAberration} objects. \textit{WavefrontSensor} objects must always provide a reference to the \textit{Source} that they are sensing.

Objects shall not be instantiated as \textit{WavefrontSensor}, but specifically as one of the two subclasses provided: \textit{ShackHartmann} or \textit{Pyramid}. These subclasses support extra data that are specific to that type of Wavefront Sensor. Support for further types may be provided in the future.

A detailed representation of the contents of this class, its subclasses and interactions with other classes can be seen in Fig.~\ref{fig:optical_sensor}. 

\subsubsection{WavefrontCorrector}
\begin{figure}[htbp]
   \begin{center}
   \includegraphics[scale=0.27]{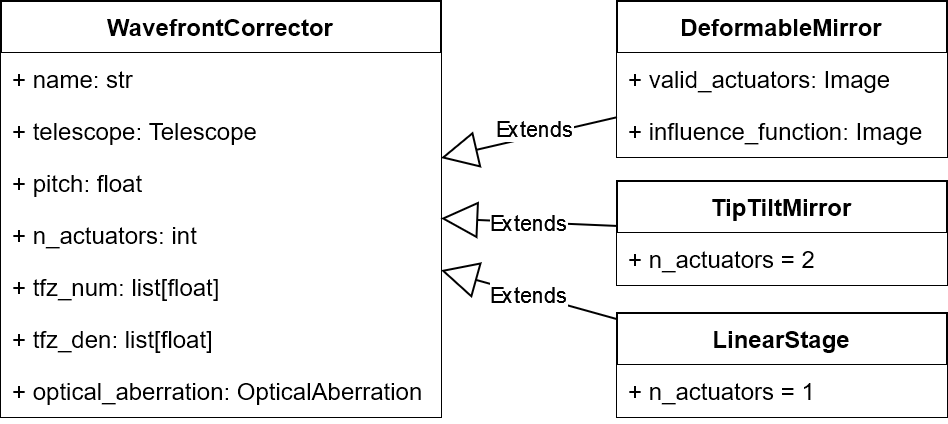}
   \end{center}
   \caption{\label{fig:wavefront_corrector} 
    UML Class Diagram for the \textit{WavefrontCorrector} class and its subclasses \textit{DeformableMirror}, \textit{TipTiltMirror} and \textit{LinearStage}.}
\end{figure}

An instance of this class characterises a set of actuators that are commanded jointly to correct wavefronts. \textit{WavefrontCorrector} objects must always provide a reference to the \textit{Telescope} that they are installed on (either a \textit{MainTelescope} or a \textit{LaserLaunchTelescope}) and may also provide a reference to related \textit{OpticalAberration} objects. Objects shall not be instantiated as \textit{WavefrontCorrector}, but specifically as one of the three subclasses provided: \textit{DeformableMirror}, \textit{TipTiltMirror} or \textit{LinearStage}. The difference in these types of Wavefront Correctors lies mostly on the number of actuators involved (a Linear Stage has a single actuator, for translation, a Tip-Tilt Mirror has two actuators, for tip and tilt, and a Deformable Mirror may have any number of actuators larger than 2). Other types of correctors may be supported in the future. A diagram for this class is provided in Fig.~\ref{fig:wavefront_corrector}.

\subsubsection{RTC}
The \textit{RTC} (Real-Time Computer) is a conceptual object that holds lookup tables (LUTs) and non-common path aberrations (NCPAs) that may be used in AO computation, as well as a set of \textit{Loop} objects.

In the context of AO, a loop may be generally understood as the process through which the RTC periodically gathers data from a certain input and calculates a resulting output data, which is usually a set of commands sent to some sort of \textit{WavefrontCorrector}. In the case of \textit{ControlLoop} objects, wavefront data (as sensed by a \textit{WavefrontSensor}) are taken as the input, from which the RTC calculates the necessary adjustments to a \textit{WavefrontCorrector}, based on interaction and control matrices. As for \textit{OffloadLoop} objects, the input is instead the set of commands received by one \textit{WavefrontCorrector}, which then results on an output which is offloaded to another \textit{WavefrontCorrector}, based on a offload matrix. \textit{Loop} objects shall be instantiated as their specific subclass, in order to properly setup the objects they coordinate.

In the real-world, an AO loop may command multiple wavefront correctors based on a single wavefront sensor, or even multiple wavefront sensors may be used to command a single wavefront corrector. While \textit{aotpy} can only instantiate one-to-one relationships, this limitation can easily be worked around by instantiating multiple \textit{Loop} objects that represent the same real-world conceptual loop\footnote[1]{For example, if a real-world AO loop uses a single wavefront sensor to command multiple wavefront correctors, in \textit{aotpy} we can create one \textit{ControlLoop} object for each wavefront corrector being commanded. These objects will all have the same \textit{WavefrontSensor} object as input, but their control matrices and commands will differ based on their respective \textit{WavefrontCorrector} object (output).}, with minimal overhead (given that the duplicated data is mostly composed of references).

\begin{figure}[htbp]
   \begin{center}
   \includegraphics[scale=0.27]{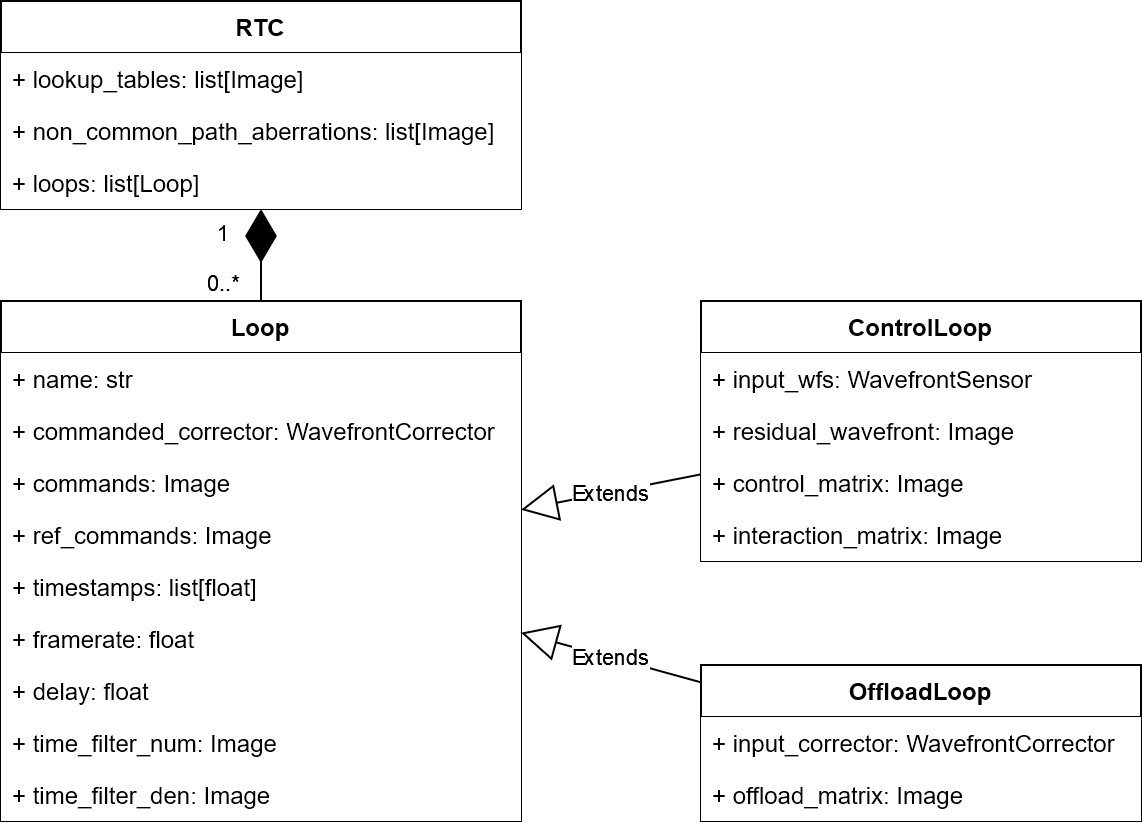}
   \end{center}
   \caption{\label{fig:rtc} 
    UML Class Diagram for the \textit{RTC} class as well as \textit{Loop} and its subclasses \textit{ControlLoop} and \textit{OffloadLoop}.}
\end{figure}

It is important to note that, since \textit{ControlLoop} objects provide references to one \textit{WavefrontSensor} object and one \textit{WavefrontCorrector}, which themselves provide references to one \textit{Source} object and one \textit{Telescope} object respectively, we can access the entire chain of objects that are related to a single \textit{ControlLoop} object via references. This property can be easily observed in the overview of the data model in Fig.~\ref{fig:UMLClassDiagram}, as all of these objects are linked together by aggregation. For a more specific view into the \textit{RTC}, \textit{Loop} and its subclasses, Fig.~\ref{fig:rtc} is provided.

\subsection{Readers and Writers}
\begin{figure}[htbp]
   \begin{center}
   \includegraphics[width=\textwidth]{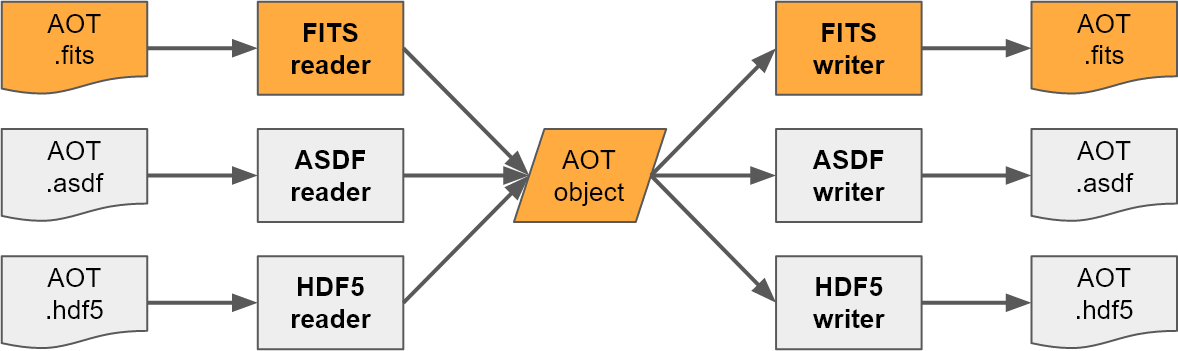}
   \end{center}
   \caption{\label{fig:readerwriter} 
    Diagram representing the functionality of the Readers and Writers in \textit{aotpy}. The FITS Reader and Writer are currently implemented, while the others are examples of file types that could be supported. The ``AOT object'' here is a \textit{AOSystem} object, which can be freely edited before being written again.}
\end{figure}

In \textit{aotpy}, the task of a Reader is to receive one AOT file as input and return a single \textit{AOSystem} object. Conversely, a Writer is able to write an \textit{AOSystem} object into an AOT file. These tools completely abstract users from the file handling process, allowing them to analyse, edit and store AO telemetry without any prior knowledge of the actual file structure and specification. It is important to note that the reading and writing process is independent from the actual source of the data or its contents, as it simply handles a standardised \textit{AOSystem} object. This means that, for AOT files, we can handle data produced by any system with a single Reader/Writer pair, for a given file type. A diagram representing the overall functionality of the Readers and Writers can be seen in Fig.~\ref{fig:readerwriter}.

Currently, only a FITS Reader and a FITS Writer are provided, but other file types (such as ASDF\cite{greenfield_asdf_2015} or HDF5\cite{hdf5}) could be supported by creating corresponding Reader/Writer pairs, without requiring any change to the data model or classes provided by \textit{aotpy} (though this is not currently planned). The FITS Reader/Writer implementation makes heavy use of Astropy\cite{astropy:2018}'s FITS File Handling package. For all of the main classes, the FITS writer essentially aggregates objects of that type and stores their data in one table per type, with the exception of \textit{Image} objects, which are stored separately as FITS image extensions. 

\subsection{Translators}
\begin{figure}[htbp]
   \begin{center}
   \includegraphics[width=\textwidth]{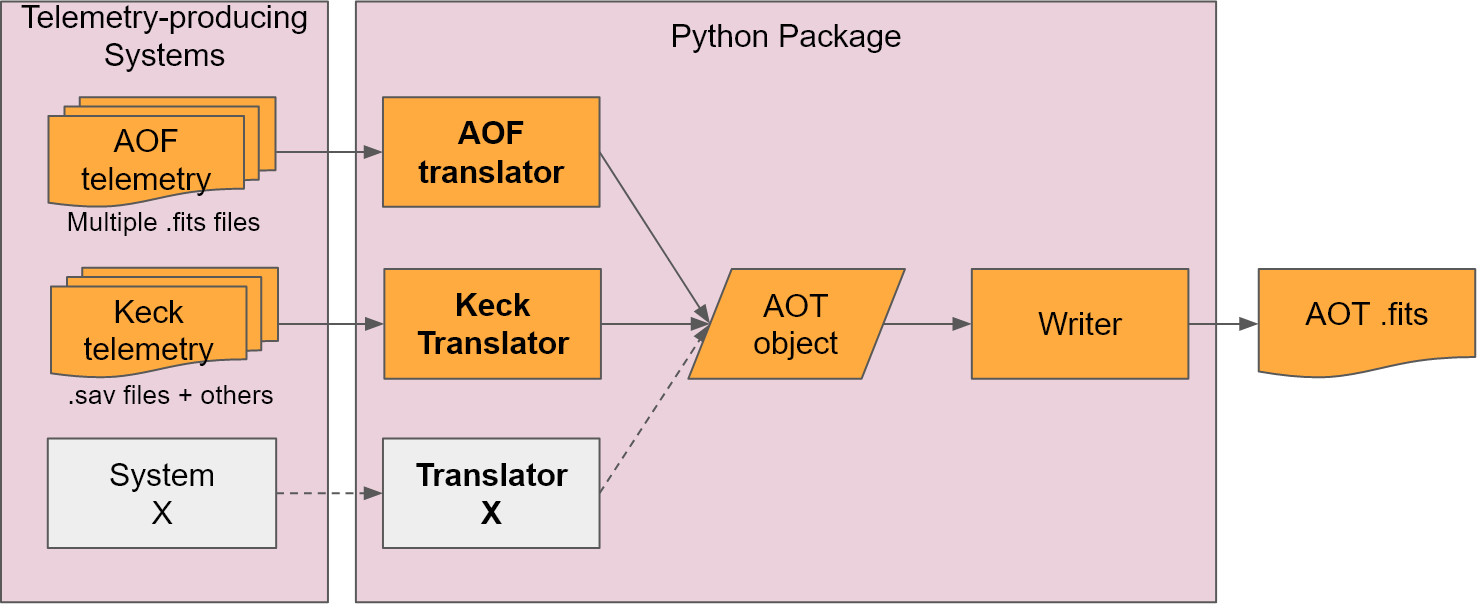}
   \end{center}
   \caption{\label{fig:translator} 
    Diagram representing the functionality of Translators in \textit{aotpy}. The translators for AOF and Keck are currently implemented, while X represents other translators that could be supported, \textit{e.g.} NAOMI or CIAO. The standard \textit{AOSystem} object could be directly used for any pipeline that accepts these objects, so writing the object to a file is purely optional.}
\end{figure}

While ideally all data-producing systems would adopt the AOT standard, in practice the current state of AO telemetry data is mostly not standardised (as previously described in Sec.~\ref{sec:intro}). This means that most data currently available cannot to be read by the Readers provided by this package. To ease the adoption of AOT, \textit{aotpy} aims to provide Translators for a set of relevant data-producing systems. For a supported data-producing system, a Translator is able to read a given set of files (produced by that system) that are related to a single observation, and then use that data to populate the corresponding fields of an \textit{AOSystem} object. This object can then be handled as any other \textit{AOSystem} object, particularly it can then be written to an AOT file, allowing for future reading of that data without requiring the user to go through the translation process again. A conceptual diagram of this feature can be seen in Fig.~\ref{fig:translator}.

Translators are highly system specific, as they require in-depth knowledge of how a given system stores its data. As such, their development requires significant collaboration with system designers/engineers familiar with that system. Therefore, \textit{aotpy} is currently only able to provide preliminary Translator support for two systems: the Adaptive Optics Facility (AOF) at ESO's VLT\cite{aof2008} and the AO system at the W. M. Keck Observatory\cite{Wizinowich_2000}. Preliminary support for other systems in ESO's VLT (specifically, NAOMI\cite{naomi2014, naomi2016} and CIAO\cite{ciao2012}) is currently under development.

AOF telemetry data is usually recorded manually and takes the form of more than 50 distinct FITS files per observation, which together contain intensity and gradient data of both of its wavefront sensors, the pixels of the NGS WFS detector, the positions of its DSM and other correctors present in its laser launch telescopes, control/offload/interaction matrices and a wide range of data regarding the characteristics of the detectors in the system. The AOF translator is able to package all of the relevant data made available in these files, along with a few configuration matrices that were sourced externally, into the standardised \textit{AOSystem} object. 

The Translator for Keck currently supports both its NGS and LGS mode, for the NIRC2 instrument. Telemetry data is provided in the form of an IDL SAVE file (for which SciPy\cite{2020SciPy-NMeth}'s \textit{readsav} routine is used), which mostly contains data regarding its wavefront sensors (specifically slopes for one or two of its WFS, depending on the mode), the commands sent to its deformable and tip-tilt mirrors and the corresponding control/interaction matrices. Relevant atmospheric data can be obtained through the Mauna Kea Weather Center seeing page\cite{lyman2021mauna}. Other relevant data such as pupil masks, NCPAs and filter characteristics has been sourced externally.

\subsection{Code Demonstrations}
In order to exemplify the usage of this package and to familiarise the reader with its workflow, in the following sections we provide three examples of common tasks that may be performed with \textit{aotpy}, each having code excerpts along with explanations of the logic behind them.
\subsubsection{Creating an AOT file from scratch}
\label{sec:1stexample}
Let's say that we have some AO telemetry data that we want to store following the AOT specification. For this demonstration we will assume that we have a system with one NGS and one LGS, each being sensed by a different Shack-Hartmann wavefront sensor (with 4 subapertures), and the RTC uses data from these WFSs to calculate commands for a single Deformable Mirror in the system (32 actuators). The data is assumed to have been recorded for 10000 frames. For simplicity, we will focus mostly on the slopes and commands data, but keep in mind that we could add large amounts of data related to physical characteristics of all the objects.

We'll start by importing all the classes that we will need for this example, as well as NumPy, which we will use to create some dummy data for demonstration purposes.
\begin{minted}{python}
import numpy as np
from aotpy import AOSystem, RTC, MainTelescope, NaturalGuideStar, LaserGuideStar,\
    ShackHartmann, ControlLoop, DeformableMirror, Image
from aotpy.fits import write_to_fits
\end{minted}
Then, we create our main telescope, as well as its single DM. 
\begin{minted}{python}
tel = MainTelescope(name="Example telescope")
cor = DeformableMirror(name="Example DM", n_actuators=32, telescope=tel)
\end{minted}
We then create an object that represents our NGS, followed by the wavefront sensor that is sensing that source. For this wavefront sensor, we provide a 10000-by-8 image, representing the vertical and horizontal slopes of the 4 subapertures during all of the recorded frames. 
\begin{minted}{python}
ngs = NaturalGuideStar(name="Example NGS")
ngs_wfs = ShackHartmann(name="WFS1", source=ngs,
                        slopes=Image(name="NGS slopes", data=np.ones((10000, 8))))
\end{minted}
After that, we do a similar process for the LGS, making sure we use the appropriate class.  
\begin{minted}{python}
lgs = LaserGuideStar(name="Example LGS")
lgs_wfs = ShackHartmann(name="WFS2", source=lgs,
                        slopes=Image(name="LGS slopes", data=np.ones((10000, 8))))
\end{minted}
Please note that, if in a certain scenario we knew that two different objects in our system were referencing the same multi-dimensional data, instead of creating two separate \textit{Image} objects we could create a single \textit{Image} object that is referenced by both of them. This would be preferable as it would avoid data duplication in the \textit{AOSystem} object and also in any resulting AOT files. That said, while in this demonstration the slopes data happens to be the exact same for both wavefront sensors, in a real-world example this would be unlikely, so we kept these as two separate \textit{Image} objects.

\begin{figure}[htbp]
   \begin{center}
   \includegraphics[scale=0.6]{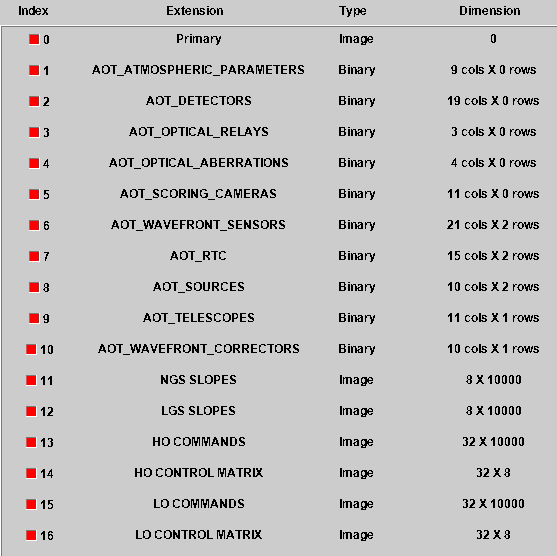}
   \end{center}
   \caption{\label{fig:fv1stexample} 
    Screenshot of the ``example.fits'' file created in Sec.~\ref{sec:1stexample} being opened with fv FITS viewer\cite{pence1997fv, pence2020fv}. As expected, we have two entries in the wavefront sensors, sources and RTC tables, while we have just one entry for the telescope and wavefront correctors tables. All the remaining tables have no entries. We also have 6 image extensions, which match the images created for the slopes, commands and control matrices.}
\end{figure}

With the previous steps completed, we now can create the control loops in the system. Let's assume we have a High Order (HO) loop that uses data from the LGS, and a Low Order (LO) loop that uses data from the NGS. For each loop, we have a 10000-by-32 image that represents the commands sent to the DM. We also have the two 8-by-32 control matrices that were used to create the commands. 
\begin{minted}{python}
lo_loop = ControlLoop(name="LO loop", input_wfs=ngs_wfs, commanded_corrector=cor,
                      commands=Image(name="LO commands", data=np.ones((10000, 32))),
                      control_matrix=Image(name='LO control matrix', data=np.ones((8, 32))))
ho_loop = ControlLoop(name="HO loop", input_wfs=lgs_wfs, commanded_corrector=cor,
                      commands=Image(name="HO commands", data=np.ones((10000, 32))),
                      control_matrix=Image(name='HO control matrix', data=np.ones((8, 32))))
\end{minted}
Then we need to create an \textit{RTC} object that contains the loops that we created.
\begin{minted}{python}
rtc = RTC(loops=[lo_loop, ho_loop])
\end{minted}
To wrap it all up into a single object, we create an \textit{AOSystem} object, which receives references to the objects that we created. 
\begin{minted}{python}
system = AOSystem(sources=[ngs, lgs],
                  telescope=tel,
                  wavefront_sensors=[ngs_wfs, lgs_wfs],
                  rtc=rtc,
                  wavefront_correctors=[cor])
\end{minted}
Finally, we write this object into a file, by providing the Writer function with the relevant path (including filename), which may be relative or absolute.
\begin{minted}{python}
write_to_fits(system, "example.fits")
\end{minted}
The structure of the resulting FITS file can be seen in Fig.~\ref{fig:fv1stexample}. Note that, while NumPy handles arrays in row-major order by default, arrays stored in FITS always use column-major order, which is why the order of the dimensions appears reversed in the figure. However, the user is completely abstracted from this conversion, as it is handled automatically by \textit{aotpy}.

\subsubsection{Reading and editing an existing AOT file}
Now that we have created an AOT file in the previous section, it can be shared with anyone that might be interested in the data. While they will be able to read the file with any FITS reader that supports the latest specification, they can also read the data through \textit{aotpy}, which provides useful abstractions and enables the user to easily edit the contents.

To exemplify this, we can start by importing the relevant Reader function and then providing it with the path to the file we want to read (in this case, the file created in the previous section).
\begin{minted}{python}
from aotpy.fits import read_from_fits
system = read_from_fits('example.fits')
\end{minted}
After the reading process is finished, \textit{system} is an \textit{AOSystem} object whose contents are completely equivalent to the \textit{system} variable that we had at the end of the previous section. This means that we are essentially picking back up from where we left off in the previous section.

This variable can now be explored in the same way as any Python object can be explored, and it is also directly editable by changing the values associated with its attributes. For example, we could specify the framerate of the LO and HO loops as such: 
\begin{minted}{python}
system.rtc.loops[0].framerate = 500
system.rtc.loops[1].framerate = 1000
\end{minted}
While in the previous step we have locally edited this variable, keep in mind that this change is not automatically reflected in the file itself, by design. In order to achieve this, one would have to use a Writer to write this edited variable into a new FITS file.

\subsubsection{Translating from non-standard files}
\begin{figure}[htbp]
   \begin{center}
   \begin{tabular}{c} 
   \includegraphics[width=0.48\textwidth]{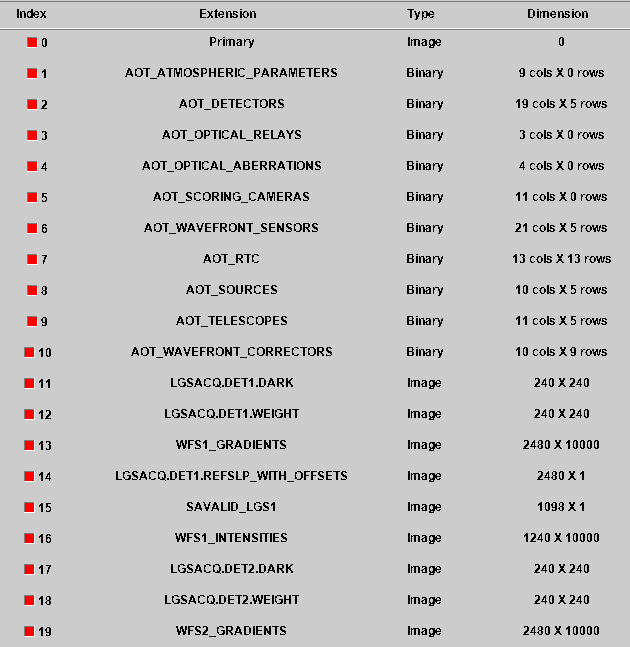}
   \includegraphics[width=0.48\textwidth]{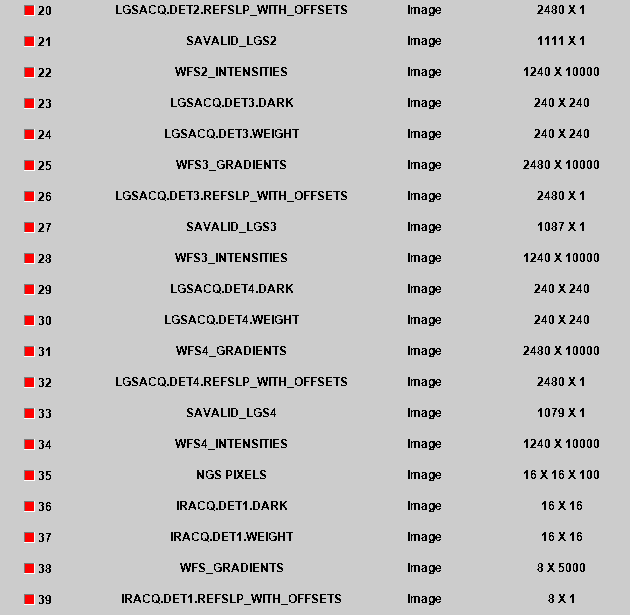}
   \end{tabular}
   \end{center}
   \caption{\label{fig:fvscreenshot} 
    Partial screenshot of fv opening an AOT file containing standardised AOF telemetry. Of 81 total extensions (HDUs) contained in this particular file, an overview of the first 40 is shown.}
\end{figure}

When a Translator for a certain dataset is available, handling that data is a simple process. The first step is to import the relevant translation method and then calling it by passing the path to the folder which contains all the relevant files. For AOF data, this could be achieved in the following way:
\begin{minted}{python}
from aotpy.translators.aof import load_from_aof
system = load_from_aof('path/to/folder')
\end{minted}
From this point on, \textit{system} is a standard \textit{AOSystem} object which can be freely edited and explored. We could, for example, write this object into an AOT FITS file:
\begin{minted}{python}
from aotpy.fits import write_to_fits
write_to_fits(system, 'aof.fits')
\end{minted}
An overview of the contents of the resulting ``aof.fits'' file can be seen in Fig.~\ref{fig:fvscreenshot}.

\section{EXAMPLE USE-CASES}
To demonstrate the usefulness of AO telemetry data, we provide an example of a scientifically relevant use-case with real-world data. Specifically, a topic of recent investigation efforts is the estimation of atmospheric turbulence parameters from Shack–Hartmann slopes\cite{andrade2019estimation}.

\begin{figure}[htbp]
   \begin{center}
   \includegraphics[width=0.7\textwidth]{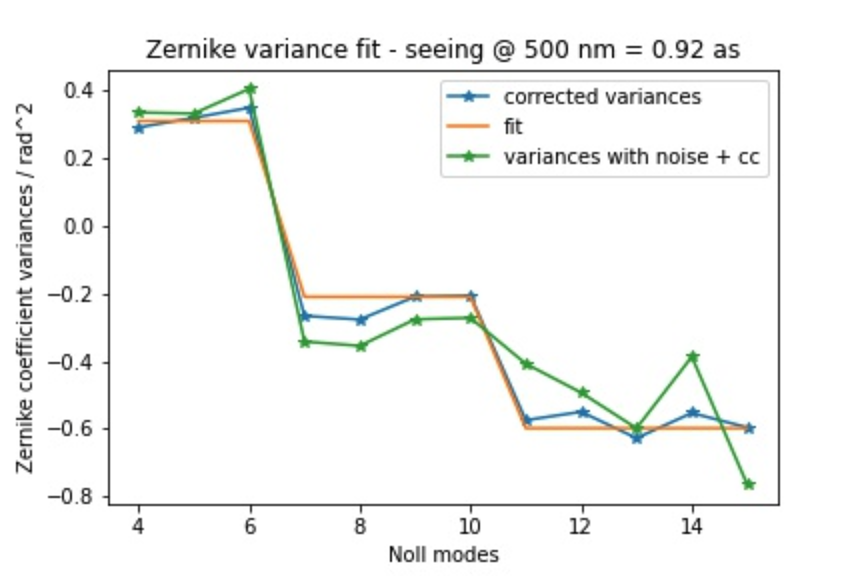}
   \end{center}
   \caption{\label{fig:turbulenceparameterestimation} 
    Turbulence estimation fit utilising telemetry data for the NAOMI system. Utilising the methods described in Ref.~\citenum{andrade2019estimation} we can obtain an estimation of the characteristic parameters of turbulence, the Fried parameter and the outer-scale, by correcting for the effects of cross-talk and aliasing inherently present in the finite sensor measurement. From the obtained fit we can estimate a $r_0$ of (11.36 $\pm$ 1.02) cm and a $L_0$ of (14.1 $\pm$ 1.5) m.}
\end{figure}

Using a work-in-progress \textit{aotpy} Translator for NAOMI data, we created an \textit{AOSystem} object that was then fed into an internal pipeline for estimating atmospheric turbulence parameters. The resulting estimation is illustrated in Fig.~\ref{fig:turbulenceparameterestimation}.

\section{FUTURE WORK}
Given the importance of community acceptance, we are developing this project following a ``bottom-up'' approach. We start by releasing the \textit{aotpy} package to the community (fully open-sourced) in its current state, as a beta version. These supporting tools allow the community to experiment with the standard and provide feedback at an earlier stage, allowing us to rapidly iterate the standard and package to fix any issues that arise and to better match the needs of the community. 

Then, in Fall 2022, we will release a paper fully describing the AOT FITS standard, which will mark the release of its version 1.0. Along with it the version 1.0 of the \textit{aotpy} package will also be released, marking the end of its beta phase. From this point forward, file backwards compatibility will be guaranteed, ensuring that a file that respects any version of the AOT standard will always be readable by the latest version \textit{aotpy}. Although, it is expected that the new versions of AOT FITS standard may be developed in the future (with expansions to meet advancements in the field and user needs), these will be accompanied by new \textit{aotpy} versions that guarantee compatibility. Specifically, we aim to adequately support future ELT-class observatories, by making any necessary updates as their requirements become clearer.

In the short term, \textit{aotpy} will provide Translator support for additional ESO systems (NAOMI and CIAO). A selection of datasets from ESO systems will be made available in the ESO Science Archive, following the AOT standard.

\acknowledgments
This project has received funding from the European Union’s Horizon 2020 research and innovation programme under grant agreements No. 730562 (OPTICON) and 101004719 (OPTICON–RadioNet Pilot).

\bibliography{report} 
\bibliographystyle{spiebib} 

\end{document}